\documentclass[twocolumn]{svjour3}
\usepackage{graphics}
\usepackage{amsmath}
\usepackage{amssymb}
\newcommand{\EQ}[3]{
  \begin{equation}
    \label{#1}
    #2
    \;#3
  \end{equation}
}

\def\hat{\mathaccent "705E\relax}

\journalname{epjd}
\begin{document}
%
\title{Time-dependent condensation of bosonic potassium}
\author{A.\,Kabelac \and G.\,Wolschin}
\institute{Institut f\"ur Theoretische Physik der Universit\"at Heidelberg, Philosophenweg 16, D-69120 Heidelberg, Germany, EU}
\date{Received: date / Revised version: date}
%
\abstract{
We calculate the time-dependent formation of Bose--Einstein condensates (BECs) in potassium va-\\pours 
based on a previously derived exactly solvable nonlinear boson diffusion equation (NBDE). Thermalization following a
 sudden energy quench from an initial temperature 
$T_\mathrm{i}$ to a final temperature $T_\mathrm{f}$ below the critical value and BEC formation are accounted for using closed-form analytical solutions of the NBDE. The time-dependent condensate fraction is compared with available $^{39}$K data for various scattering lengths. 
%
} 
\maketitle
\section{Introduction}
\label{intro}
The time-dependent mathematical modelling of nonequilibrium phenomena  is an outstanding problem in physics:
How does an isolated quantum many-body system thermalize that is originally far from equilibrium?
This problem is particularly relevant and interesting for ultracold Bose gases, where in the course of thermalization a transition to a Bose--Einstein condensate (BEC) can occur below a critical temperature $T_\mathrm{c}$ which it is not yet fully understood \cite{gli21}. 

Ever since and even before the discovery of BEC in alkali atoms, numerical nonequilibrium theories have been developed
to account for the time-dependence of thermalization and condensate formation. Predictions such as \cite{gz97} for sodium and rubidium
had been made. For $^{23}$Na, some of these theoretical results for the time-dependent condensate growth \cite{bzs00} have been compared with experimental data \cite{miesner_bosonic_1998}.

 For $^{87}$Rb, data have been compared to a numerical model based on quantum kinetic theory \cite{kdg02}. Here, cooling into the quantum-degenerate regime was achieved by continuous evaporation for a duration of several seconds, whereas in the sodium experiment a short radio-frequency pulse was applied to remove high-energy atoms.

Only recently, new experimental results for potassium have become available with a sufficiently detailed time resolution: Thermalization and condensate formation is investigated in
a homogeneous 3D Bose gas of $^{39}$K with tunable interactions and near-perfect isolation in a cylindrical optical box \cite{gli21}.

As a transparent and analytically solvable model for the time-dependence of thermalization and condensate formation of bosonic systems, a nonlinear boson diffusion equation (NBDE) has been derived in Ref.\,\cite{gw18}, applied to ultracold atoms \cite{gw18a,gw20}, and solved exactly with the necessary boundary 
conditions at the singularity \cite{rgw20}. We have used the model for the calculation of the time-dependent condensate fraction in $^{23}$Na \cite{sgw21}, where we have  tested it successfully against the historical MIT data \cite{miesner_bosonic_1998}. In the present work, we perform a related calculation for $^{39}$K to compare with the newly available Cambridge data for tunable scattering lengths \cite{gli21}.


\section{The model}
First we briefly review our previously developed \cite{gw18,gw18a,gw20,rgw20,sgw21} nonlinear model that we use in this work to compute the nonequilibrium evolution following a sudden energy quench, and time-dependent condensate fractions in equilibrating Bose gases of ultracold atoms with a focus on a comparison with recent data \cite{gli21} for $^{39}$K.
The cold-atom vapour in a trap is described as a time-dependent mean field with a collision term and we consider only $s$-wave scattering 
with the corresponding scattering lengths. The $N$-body density operator ${\hat{\rho}_N(t)}$ is composed of $N$ single-particle wave functions of the atoms which are solutions of the time-dependent Hartree-Fock equations plus a time-irreversible collision term $\hat{K}_{N}(t)$ that causes the system to thermalize through random two-body collisions 
($\hbar=c=1$)
\EQ{}{
i\,\frac{ \partial\hat{\rho}_N(t)}{\partial t} = \big[\hat{H}_\mathrm{HF}(t),\hat{\rho}_{N}(t)\big] + i \hat{K}_{N}(t)
\label{eq1}}{}
with the self-consistent Hartree-Fock mean field $\hat{H}_\mathrm{HF}(t)$ of the atoms in a trap that provides an external potential.

The full many-body problem is reduced to the one-body level in an approximate version for the ensemble-averaged single-particle density operator 
 $\bar{\rho}_1(t)$. Its diagonal elements represent the 
probability for a particle to be in a state $|\alpha\rangle$ with energy $\epsilon_\alpha$
\EQ{}
{\big(\bar{\rho}_1(t)\big)_{\alpha,\alpha} = n(\epsilon_\alpha, t)\equiv n_\alpha(\epsilon,t)}
{.}
The total number of particles is $N=\sum_\alpha n_\alpha$, and we neglect here
the off-diagonal terms of the density matrix. As discussed in Refs.\,\cite{gw18,gw20,sgw21}, the occupation-number distribution $n_\alpha(\epsilon,t)$ in a finite Bose system obeys a Boltzmann-like collision term, where the distribution function depends only on energy and time. 
A condition for such a reduction to $1+1$ dimensions is spatial and momentum isotropy. The corresponding assumption of sufficient ergodicity has been widely discussed in the literature  \cite{snowo89,kss92,setk95,lrw96,hwc97,jgz97}. For the thermal cloud of cold atoms surrounding a Bose--Einstein condensate (BEC), it is expected to be essentially fulfilled, even though the condensate in a trap is spatially anisotropic. Different spatial dimensions enter our present formulation through the density of states, which depends also on the type of confinement. The model calculations in this work are for a 3D system, and in line with the $^{39}$K experiment \cite{gli21}, we investigate results for the density of states of bosonic atoms confined in a cylindrical optical box.

The collision term for the single-particle 
occupation-number distribution of the energy eigenstates 
$\epsilon_\alpha$, $n\equiv n_\alpha \equiv \langle n(\epsilon_\alpha,t)\rangle$, can be transformed \cite{gw18} into a nonlinear partial differential equation   
 \begin{equation}
\frac{\partial n}{\partial t}=-\frac{\partial}{\partial\epsilon}\Bigl[v\,n\,(1+n)+n\frac{\partial D}{\partial \epsilon}\Bigr]+\frac{\partial^2}{\partial\epsilon^2}\bigl[D\,n\bigr]\,.
 \label{boseq}
\end{equation}
In this nonlinear boson diffusion equation (NBDE), the drift term $v(\epsilon,t)$ accounts for dissipative effects, whereas $D(\epsilon,t)$ mediates diffusion of particles in the energy space: It accounts for the broadening of the distribution functions, and in particular, for the softening of the sharp cut at  $\epsilon=\epsilon_\mathrm{i}$ that signifies the quench, as well as for the diffusion of particles into the condensed state.
The many-body physics is contained in these
transport coefficients, which depend on energy, time, and the second moment of the interaction. 

It has been shown \cite{sgw21} that the stationary solution $n_\infty(\epsilon)$ of the NBDE for $t\rightarrow\infty$ equals the Bose--Einstein equilibrium distribution $n_\mathrm{eq}(\epsilon)$
\begin{equation}
n_\infty(\epsilon)=n_\mathrm{eq}(\epsilon)=\frac{1}{e^{(\epsilon-\mu)/T}-1}\,,
 \label{Bose--Einstein}
\end{equation}
provided the ratio $v/D$ has no energy dependence, requiring
$\lim_{t\rightarrow \infty}[-v(\epsilon,t)/D(\epsilon,t)] \equiv 1/T$. 
The chemical potential is $\mu\leq0$ in a finite Bose system.

In the limit of energy-independent transport coefficients, the nonlinear boson diffusion equation  
for the occupation-number distribution $n(\epsilon,t)$
becomes
\begin{equation}
\frac{\partial n}{\partial t}=-v\,\frac{\partial}{\partial\epsilon}\Bigl[n\,(1+n)\Bigr]+D\,\frac{\partial^2n}{\partial\epsilon^2}\,.
 \label{bose}
\end{equation}
Again, the thermal equilibrium distribution $n_\mathrm{eq}$ is a stationary solution
with $\mu\leq0$ and $T=-D/v$. 

The NBDE with constant transport coefficients 
preserves the essential features of Bose--Einstein
statistics that are contained in the bosonic Boltzmann equation. It is one of the few nonlinear partial differential equations with a clear physical meaning that can be solved exactly through a nonlinear transformation, as outlined in  Refs.\,\cite{gw18,gw18a,gw20}. The resulting solution is
\begin{eqnarray}
    n(\epsilon,t) = T \frac{\partial}{\partial\epsilon}\ln{\mathcal{Z}(\epsilon,t)} -\frac{1}{2}= T\frac{1}{\mathcal{Z}} \frac{\partial\mathcal{Z}}{\partial\epsilon} -\frac{1}{2}\,,
    \label{net} 
    \end{eqnarray}
 where the time-dependent partition function ${\mathcal{Z}(\epsilon,t)}$ obeys a linear diffusion equation which has a Gaussian Green's function $G(\epsilon,x,t)$.
The partition function can then be written as an integral over the Green's function and an exponential function $F(x)$ 
     \begin{eqnarray}
    \mathcal{Z}(\epsilon,t)= \int_{-\infty}^{+\infty} G(\epsilon,x,t)\,F(x)\,\mathrm{d}x\,.
    \label{eq:partitionfunctionZ}
    \end{eqnarray}
    $F(x)$ depends on 
 the initial occupation-number distribution $n_\mathrm{i}$
according to
 \begin{eqnarray}
    F(x) = \exp\Bigl[ -\frac{1}{2D}\bigl( v x+2v \int_0^x n_{\mathrm{i}}(y)\,\mathrm{d}y \bigr) \Bigr]\,.
       \label{ini}
\end{eqnarray}
The definite integral over the initial conditions taken at the lower limit in Eq.\,(\ref{ini}) drops out in the calculation of $n(\epsilon,t)$ and can be replaced \cite{rgw20} by the indefinite integral
with the integration constant set to zero without affecting the accuracy of the calculation. 
These modifications allow us to compute the partition function and the overall solution for the occupation-number distribution function Eq.\,(\ref{net}) analytically.

With the free Green's function $G\equiv G_\mathrm{free}$ of the linear diffusion equation, the physically correct solution with the Bose--Einstein equilibrium limit is attained in the UV region, but not in the IR \cite{gw18}. To solve this problem, one has to consider the boundary conditions at the singularity $\epsilon = \mu \leq 0$ \cite{gw20}. They can be expressed as
 \(\lim_{\epsilon \downarrow \mu} n(\epsilon,t) = \infty\) \,$\forall$ \(t\). One obtains a vanishing partition function at the boundary \( \mathcal{Z} (\mu,t) = 0\), and the energy range is restricted to  $\epsilon \ge \mu$. This requires a new Green's function \cite{gw20} 
that equals zero at \(\epsilon = \mu\) $\forall \,t$. It can be written as
\begin{eqnarray}
    {G}_\mathrm{b} (\epsilon,x,t) = G_\mathrm{free}(\epsilon - \mu,x,t) - G_\mathrm{free}(\epsilon - \mu,-x,t)\,,
    \label{eq:newGreens}
\end{eqnarray}
    and the partition function with this boundary condition becomes
     \begin{eqnarray}
    \mathcal{Z}_\mathrm{b} (\epsilon,t)= \int_{\mu}^{+\infty} G_\mathrm{b} (\epsilon,x,t)\,F(x)\,\mathrm{d}x\,,
    \label{eq:partitionfunctionZ}
    \end{eqnarray}
which is equivalent to 
$\int_{0}^{+\infty} G_\mathrm{b} (\epsilon,x,t)\,F(x+\mu)\,\mathrm{d}x$:
The function $F$ remains unaltered with respect to Eq.\,(\ref{ini}), but its argument is shifted by the chemical potential.

For fixed chemical potential $\mu$ and an initial temperature $T_\mathrm{i}$ that differs from the final temperature $T_\mathrm{f} = - D/v$ as in an energy quench, we have solved the combined initial- and boundary value problem exactly in Ref.\,\cite{rgw20} using the above nonlinear transformation \cite{gw20} from Eq.\,(\ref{net}). Expressions of the form $[1-\exp{(-z/T_\mathrm{i})}]^{T_\mathrm{i}/T_\mathrm{f}}$ that appear in $F(x)$ and thus, in the partition function Eq.\,(\ref{eq:partitionfunctionZ}), can be reformulated using
the generalized binomial theorem
\begin{equation}
(1+x)^s=\sum_{k=0}^{\infty} \binom{s}{k} x^k\,,
\end{equation}
which results in an infinite series expansion for the time-dependent partition function (and also for its derivative), 
\begin{equation}
{\mathcal{Z}}(\epsilon,t) = \sqrt{4 D t} \, \exp\Bigl(-\frac{\mu}{2 T_{\mathrm{f}}}\Bigr) \sum_{k=0}^{\infty} \binom{\frac{T_{\mathrm{i}}}{T_{\mathrm{f}}}}{k} \left( -1 \right)^k \times
f_k^{\text{T}_\text{i},\text{T}_\text{f}} \,(\epsilon,t)\,.
\label{partfct}
\end{equation}
The analytical expressions for $f_k^{\text{T}_\text{i},\text{T}_\text{f}} \,(\epsilon,t)$ are combinations of exponentials and error functions, they are given explicitly in
 Ref.\,\cite{rgw20}. We can now compute the distribution function directly from Eq.\,(\ref{net}). The convergence of the solutions has been studied in  Ref.\,\cite{sgw21}, indicating that 
-- depending on the system and the specific observable -- a result with $k_\mathrm{max}=10-40$ expansion coefficients is indistinguishable from the exact solution with $k_\mathrm{max}=\infty$. It has also been shown that the series terminates in case $T_\mathrm{i}/T_\mathrm{f}$ is an integer, and we shall explicitly use that property in this work, where the exact solution is already obtained for
$k_\mathrm{max}=4$. A brief summary of the nonlinear diffusion model for both, boson and fermion systems at low and high energies and temperatures is presented in Ref.\,\cite{gw22}.

\section{Energy quench and thermalization}
%
\begin{figure}
\resizebox{0.48\textwidth}{!}{%
  \includegraphics{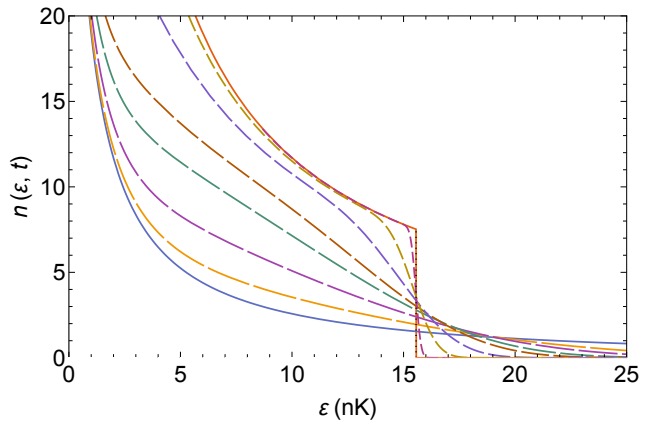}
}
\caption{Nonequilibrium evolution of quenched $^{39}$K vapour calculated from the analytical solution of the nonlinear boson diffusion equation (NBDE, \cite{gw18,gw20}) for fixed chemical potential.
The initial state is a Bose--Einstein distribution with $T_\mathrm{i} = 130$\,nK, $\mu_\mathrm{i}=-0.67$\,nK truncated at $\epsilon_{i}$\,=\,15.56\,nK (see  Eq.\,(\ref{ni0})), upper solid curve.
	The transport coefficients are $D = 0.08$\,(nK)$^2$/ms, $v = -0.00246$\,nK/ms. The final temperature is $T_\mathrm{f}=-D/v = 32.5$\,nK (lower solid curve).
	The time evolution of the single-particle occupation-number distributions is shown at  $t = 0.2, 4, 20, 60, 100, 200$, and 400\,ms (increasing dash lengths). 
}
\label{fig1}       
\end{figure}
\begin{figure}
\resizebox{0.48\textwidth}{!}{%
  \includegraphics{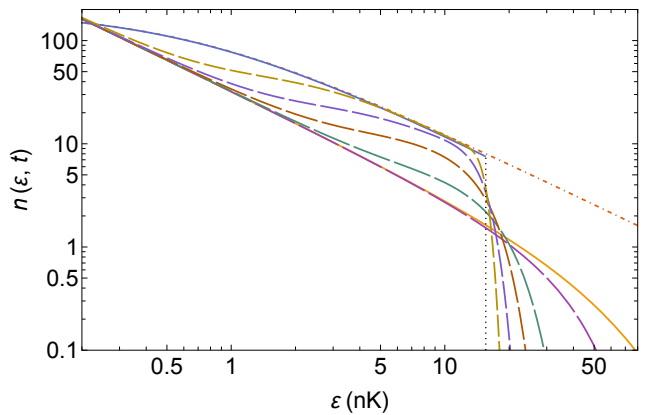}
}
\caption{Nonequilibrium evolution of $^{39}$K vapour  based on the NBDE as in Fig.\,\ref{fig1}, but using $\mu=0$ and a double-log-scale. Thermalization in the UV tail is significantly slower than below the evaporative cut and in particular, in the IR. Timesteps are $t = 8, 30, 100, 300$, and 4000\,ms. The dot-dashed line is the Rayleigh--Jeans power law in the initial distribution.}\label{fig2}       
\end{figure}


In the $^{39}$K experiment of Ref.\,\cite{gli21} with tunable interactions, a far-from-equilibrium atomic cloud is created by removing $77\%$ of the atoms and $\simeq97.5\%$ of the total energy through an energy quench below the critical temperature, thus setting the stage for subsequent condensate formation. This quench corresponds to cutting off the atoms with energy 
$\epsilon>\epsilon_\mathrm{i}$ according to
\begin{equation}
n_\mathrm{i}(\epsilon)=\frac{1}{\exp{(\frac{\epsilon-\mu_\mathrm{i}}{T_\mathrm{i}})}-1}\theta(1-\epsilon/\epsilon_\mathrm{i})\,.
\label{ni0}
\end{equation}
Hence, the number of atoms just after the quench 
becomes
with the 3D density of states $g(\epsilon)\propto \sqrt{\epsilon}$ in a box 
\begin{eqnarray}
N_\mathrm{i}\propto&\int_0^{\epsilon_\mathrm{i} }\frac{{\sqrt{\epsilon}}}{\exp{(\frac{\epsilon-\mu_\mathrm{i}}{T_\mathrm{i}})}-1}\,\mathrm{d}{\epsilon}\nonumber\\
&=0.23\,\int_0^{\infty}\frac{\sqrt{\epsilon}}{\exp{(\frac{\epsilon-\mu_\mathrm{i}}{T_\mathrm{i}})}-1}\,\mathrm{d}{\epsilon}\,.
\label{ni}
\end{eqnarray}
In agreement with experiment \cite{gli21}, it is kept constant in the subsequent thermalization and condensate formation process.
The equilibrium value of the ensuing time-dependent condensate fraction for $t\rightarrow\infty$ can then be calculated from the ratio of the difference in particle content
of the initial nonequilibrium distribution at finite $\mu_\mathrm{i}<0$ 
 minus the final equilibrium Bose--Einstein distribution with temperature $T_\mathrm{f}$ and vanishing chemical potential,  
 divided by particle content of the initial distribution. The result is\\
\begin{eqnarray}
\frac{N_\mathrm{c}^\mathrm{eq}}{N_\mathrm{i}}=&1-\int_0^{\infty} g(\epsilon)n_\mathrm{eq}(\epsilon)\mathrm{d}{\epsilon}\Big{[}\int_0^{\epsilon_\mathrm{i}} g(\epsilon)n_\mathrm{i}(\epsilon)\mathrm{d}{\epsilon}\Big{]}^{-1}\nonumber\qquad\qquad\\
=&1-T_\mathrm{f}^{3/2}\zeta(3/2)\Gamma(3/2)\Big{[}\int_0^{\epsilon_\mathrm{i} }\frac{\sqrt{\epsilon}}{\exp{(\frac{\epsilon-\mu_\mathrm{i}}{T_\mathrm{i}})}-1}\,\mathrm{d}{\epsilon}\Big{]}^{-1}
\label{nceq}
\end{eqnarray}\\
where the integration over the initial thermal distribution has an upper limit of $\epsilon=\epsilon_\mathrm{i}$, according to the cut due to the quench,
$\theta\,(1-\epsilon/\epsilon_\mathrm{i})$. The measured equilibrium condensate fraction \cite{gli21} is $N_\mathrm{c}^\mathrm{eq}/N_\mathrm{i}=40(5)\,\%$ for all scattering lengths that have been investigated, $a=(100-800)\,a_\infty$ with the Bohr radius $a_\infty$.

The above Eqs.\,(\ref{ni}),(\ref{nceq}) provide two conditions to calculate the initial value of the chemical potential 
$\mu_\mathrm{i}$, and the cut $\epsilon=\epsilon_\mathrm{i}$ that defines the quench. Inserting Eq.\,(\ref{ni}) into Eq.\,(\ref{nceq}), an implicit equation for the initial chemical potential $\mu=\mu_\mathrm{i}$ that does not depend on $\epsilon_\mathrm{i}$ is obtained, and with this value, the cut $\epsilon=\epsilon_\mathrm{i}$ can directly be computed from Eq.\,(\ref{ni}).

In the $^{39}$K-experiment, the initial temperature before the quench is $T_\mathrm{i}=130$\,nK, and the final temperature $T_\mathrm{f}=32(2)$ nK in accordance with energy- and particle-number conservation \cite{gli21}. In view of the fact  that these numbers as well as the experimental equilibrium condensate fraction of $40(5)$\,\% are approximate, we use a value of 
$T_\mathrm{f}\equiv32.5$ nK in our model calculation. This considerably simplifies the analytical calculations, because the value of $T_\mathrm{i}/T_\mathrm{f}$ is an integer, such that the series expansion of the exact analytical solution of the NBDE terminates already at $k_\mathrm{max}=4$. (A calculation with $T_\mathrm{f}=32.0$\,nK and $k_\mathrm{max}=40$ yields results which are
hardly distinguishable).

With these values for the initial and final temperatures, together with a removal rate of $77\%$ of the atoms in the quench and an equilibrium condensate fraction of $40\%$,
we obtain the resulting values for the initial chemical potential and the cooling cut\footnote{removing 97.5\% of the energy \cite{gli21} requires a slightly larger value of the cut} as
\begin{equation}
\mu_\mathrm{i}=-0.67\,\mathrm{nK},\hspace{.4cm}\epsilon_\mathrm{i}=15.56\, \mathrm{nK}\,.
\end{equation}
At and below the cut, the average initial occupation is $n_\mathrm{i}\,(\epsilon\le \epsilon_\mathrm{i})\ge7.52$. The nonequilibrium system is highly overoccupied and will move quickly into the condensed state to remove the surplus particles from the cloud.

In the simplified model with constant transport coefficients, the values of $v$ and $D$ that are required for the nonequilibrium calculation  are related to the equilibrium temperature $T_\mathrm{f}=-D/v$ and the local thermalization time $\tau_\mathrm{eq}$. The first relation is a consequence of the equality of the stationary solution of the nonlinear diffusion equation with the Bose--Einstein distribution as discussed above, whereas the thermalization time has been derived from an asymptotic expansion of the error functions in the analytical NBDE solutions \cite{gw18} for $\theta$-function initial conditions at the cut $\epsilon=\epsilon_\mathrm{i}$ as $\tau_\mathrm{eq}\equiv\tau_\infty(\epsilon=\epsilon_\mathrm{i})=4D/(9v^2)$. For a general initial distribution with a cut at $\epsilon=\epsilon_\mathrm{i}$, it can be written as
\begin{equation}
\tau_\mathrm{eq}\equiv\tau_\infty(\epsilon=\epsilon_\mathrm{i})\simeq f\,D/v^2\,,
\end{equation}
with $f=4$ for fermions \cite{gw18} and $f=4/9$ for bosons in case of initial theta-function distributions. The analytical result for a truncated Bose--Einstein initial distribution has not been derived yet. It will be considerably shorter than the one for a theta-function distribution, because the shape of the initial distribution is much closer to the final BE-result than a box distribution. Here we take $f=0.045$ to compute the transport coefficients in order to be consistent with the thermalization time of $\tau_\mathrm{eq}^\mathrm{exp}\simeq600$\,ms estimated from the data  \cite{gli21}  for $a=140\,a_\infty$.  

With a final temperature $T_\mathrm{f}=-D/v=32.5$\,nK
and a condensate-formation time $\tau_\mathrm{eq}=f\,D/v^2\simeq 600$ ms for a scattering length of $a=140\,a_\infty$, the transport coefficients that are required for the analytical solution of the NBDE are thus  obtained as \\
\begin{eqnarray}
D=&f\,T_\mathrm{f}^2/\tau_\mathrm{eq}\simeq 0.08\,\mathrm{(nK)}^2/\mathrm{ms}\,,\nonumber\\\\\nonumber
v=&-f\,T_\mathrm{f}/\tau_\mathrm{eq}\simeq-0.00246\,\mathrm{nK/ms}\,.
\end{eqnarray}\\
The nonequilibrium calculation can now be performed with the  above values for the drift and diffusion coefficients and fixed chemical potential.
First we compute the distribution functions for quenching the distribution from $T_\mathrm{i}$ to $T_\mathrm{f}$ according to
Eq.\,(\ref{net}) with the time-dependent partition function $\mathcal{Z}$ from Eq.\,(\ref{partfct}). 
The results for thermalization in $^{39}$K are shown in Figs.\,\ref{fig1}, \ref{fig2}.

Analogous analytical solutions -- albeit for different parameters -- were shown in Fig.\,2 of Ref.\,\cite{sgw21} to agree precisely with numerical Matlab-solutions of the NBDE using appropriate boundary conditions and hence, we can trust that the results are correct.

Fig.\,\ref{fig1} shows the time-dependent thermalization from the initial nonequilibrium distribution that is produced via quenching at $\epsilon_\mathrm{i}=15.56$ nK to the final Bose--Einstein equilibrium distribution with $T_\mathrm{f}=32.5$ nK and fixed chemical potential. Using the above transport coefficients, results are shown for seven timesteps at $t = 0.2, 4, 20, 60,$ $100, 200$, and $400$\, ms with increasing dash lengths. Thermal equilibrium in the cloud is achieved quickly at $t<100$\,ms in the infrared, somewhat more time is needed at intermediate energies, and much more time in the ultraviolet. 
The equilibration in the UV thermal tail is more clearly visible in
Fig.\,\ref{fig2}, which is a double-logarithmic plot with an additional timestep at 4000\,ms.
Thermalization in the cloud below the cut is seen to be essentially completed at the equilibration time $\tau_\mathrm{eq}$,
but it takes much longer in the far Maxwell--Boltzmann tail. The Rayleigh--Jeans power law in the initial distribution is clearly visible, and also power-law behaviour of the subsequent nonequilibrium evolution in limited energy intervals.

\section{Time-dependent condensate formation} 
With the analytical NBDE solutions for a given chemical potential $\mu$, we can now proceed to calculate the time-dependent transition to the condensate with the additional condition of particle-number conservation at each timestep. This requires a time-dependent chemical potential that eventually approaches zero, $\mu(t)\rightarrow 0$; from then on, particles start moving into the condensed state.
\begin{table}
\begin{center}
\caption{Transport coefficients, initiation and equilibration times for BEC formation in $^{39}$K}($T_\mathrm{i}=130$ nK, $T_\mathrm{f}=-D/v=32.5$ nK)
\vspace{.2cm}
\label{tab1}       
\begin{tabular}{rrrrr}
\hline\noalign{\smallskip}
$a\,(a_\infty)$&$D$\,(nK$^2$/ms)&$v$\,(nK/ms) &$\tau_\mathrm{ini}$\,(ms)&$\tau_\mathrm{eq}$\,(ms)\\
\noalign{\smallskip}\hline\noalign{\smallskip}
140 &0.08&$ -0.00246$&130&600 \\ 
280 &0.16&$-0.00492$&65&300\\
400&0.229&$-0.00705$&46&210\\
800 &0.457 &$-0.01406$&23&105 \\
\noalign{\smallskip}\hline
\end{tabular}
\end{center}
\end{table}
\begin{figure}
\resizebox{0.49\textwidth}{!}{%
  \includegraphics{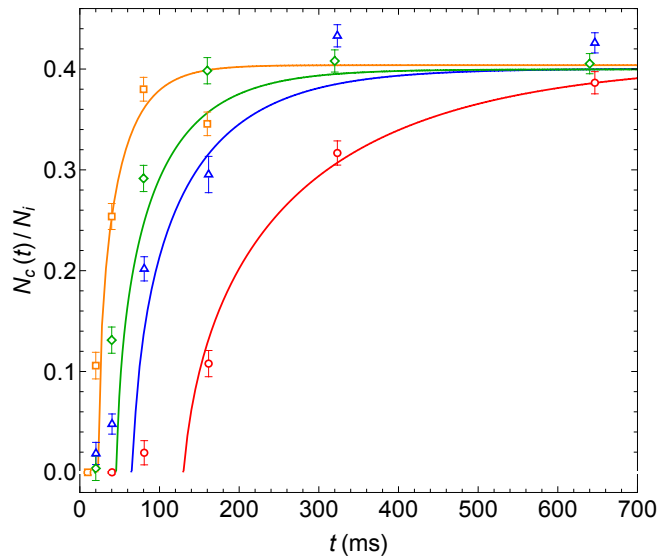}
}
\caption{Time-dependent condensate fraction for an equilibrating 3D Bose gas of $^{39}$K in an optical box for a single energy quench from $T_\mathrm{i} = 130$ nK to $T_\mathrm{f} = 32.5$ nK as calculated from the analytical solutions of the NBDE (solid curves) with $\mu_\mathrm{i} = -0.67$ nK, $\epsilon_\mathrm{i} = 15.56$ nK. Results are shown for scattering lengths $a/a_\infty=$140 (circles), 280 (triangles), 400 (diamonds) and 800 (squares), plotted with Cambridge data \cite{gli21}. The transport coefficients and time scales are given in table 1.}
\label{fig3}       
\end{figure}
The particle number in the thermal cloud at any time $t$ is given by 
\begin{equation}
N_\mathrm{th}(t)=\int_0^\infty g(\epsilon)\,\tilde{n}(\epsilon,t)\,\mathrm{d}\epsilon\,,
\label{Nth}
\end{equation}
with $\tilde{n}(\epsilon,t)$ the distribution function calculated using the value of $\mu\equiv\mu(t)\le0$ that is computed at each timestep such that the particle number is conserved, $N_\mathrm{th}(t)=N_\mathrm{i}\,\forall t$. 

Once $\mu(t)$ has reached the value zero, condensation starts, and and any further difference between the initial and the actual particle number is attributed to the condensate,
\begin{equation}
N_\mathrm{i}-N_\mathrm{th}(t)=N_\mathrm{c}(t)\,.
\end{equation}
The definition of a time-dependent chemical potential is debatable, as the distribution function for $\mu_\mathrm{i}\le\mu(t)<0$ does not have the equilibrium form.
However, for $\mu(t)<0$ we do not use the calculation to compute physical observables. It is only when equilibrium at $\mu(t)=0$ is reached and the phase transition to the condensate occurs that we start calculating the time-dependent condensate fraction from the NBDE solution plus the condition of particle-number conservation.
 Although the transport equation cannot explicitly treat the build\-up of coherence in the condensate because it does not consider phases, the nonequili\-brium-statistical approach together with the condition of particle-number conservation thus allows to properly account for the time-dependent number of particles that move into the condensed state.
 
To evaluate the integral in Eq.\,(\ref{Nth}) for the particle number in the thermal cloud once $\mu=0$ has been reached at $t=\tau_\mathrm{ini}$, no analytical solution is presently known for 
finite time $t<\infty$, but it can be evaluated numerically.
The resulting time evolution of condensate formation in $^{39}$K is calculated first with the transport parameters $D, v$ of table\,\ref{tab1} for a scattering length
The initial chemical potential is $\mu_\mathrm{i}=-0.67$\,nK and it increases with time towards zero, when condensation begins. We have tested \cite{sgw21} that no condensate forms if $T_\mathrm{f}$ at the critical number density 
$n_\mathrm{c}$ would remain above the critical temperature $T_\mathrm{c}$ for bosonic atoms of mass $m$,
\begin{equation}
T_\mathrm{c}=\frac{2\pi}{m}\left(\frac{n_\mathrm{c}}{\zeta(\frac{3}{2})}\right)^{2/3}\,.
\end{equation}
This expression is a consequence of equilibrium-statistical mechanics -- at the critical temperature, the chemical potential becomes zero. It implies that sufficient coherence for the phase transition to the condensate is achieved when the interparticle distance becomes equal to the thermal de Broglie wavelength. 
Since the Bose--Einstein equilibrium distribution is equal to the stationary limit of the NBDE for  $t\rightarrow\infty$, the critical temperature is an integral part of our nonequilibrium model, just as it is part of equilibrium statistics. Accordingly, the equilibration time 
$\tau_\mathrm{eq}$ is equivalent to the formation time of the thermalized condensate.

We can now compare the NBDE-solutions and the ensuing time-dependent (quasi-)condensate fractions
with the data \cite{gli21} for various scattering lengths, and here we focus on $a/a_\infty=140, 280, 400$ and 800. With both transport coefficients $D, v$ scaled linearly with the scattering lengths (table\,\ref{tab1}), the results are displayed in Fig.\,\ref{fig3}. The values of the transport coefficients have not been further optimzed with respect to the Cambridge data at the individual scattering lengths. A related log-log-plot is shown in Ref.\,\cite{gw22},
where the nonlinear diffusion equation is discussed in a more general context of thermalization at low and high energies as well as temperatures for both, bosonic and fermionic systems.

Condensation starts at the initiation time $\tau_\mathrm{ini}$, and rises according to the condensate-formation time 
$\tau_\mathrm{eq}$ (table\,\ref{tab1}) towards the equilibrium value $N_\mathrm{c}^\mathrm{eq}/N_\mathrm{i}$.  The initiation time required  from the data is larger than the calculated time to reach $\mu=0$ due to the assumption of constant transport coefficients, but considerably smaller than the condensate-formation time, $\tau_\mathrm{ini}\ll \tau_\mathrm{eq}$. These calculations are performed with our \texttt{C++} code that is based on the exact analytical NBDE-solutions for $n(\epsilon,t)$ from Eq.\,(\ref{net}) with $k_\mathrm{max}=4$ in the series expansion, and uses numerical integration methods to compute
 $\mu(t)$ and $N_\mathrm{c}(t)/N_\mathrm{i}$.

The resulting thermalization timescales are seen to be longest for the shortest scattering length $a=140\,a_\infty$, and decrease towards larger values of $a/a_\infty=280, 400,\\ 800$.
As is evident in particular for $a=140\,a_\infty$, small deviations in the short-time behaviour may be due to our basic assumption of constant transport coefficients, which causes condensate formation to start with full strength and not gradually. This feature appears to be absent in the $^{87}$Rb data \cite{kdg02}, but it had also been observed when comparing with $^{23}$Na data \cite{sgw21}. As an improvement, one would have to solve the full NBDE Eq.\,(\ref{boseq}) numerically with time- and energy-dependent transport coefficients.

With constant transport coefficients, we confirm with-in the nonlinear diffusion model what has already been shown in the experimental work \cite{gli21} based solely on their data: The time-dependent condensate fractions for different scattering lengths fall on a single universal curve when plotted as functions of $t'=ta/(300\,a_\infty)$, before reaching the equilibrium limit for $t'\rightarrow\infty$. 

The $1/a$-dependence of the timescales -- including that of the initiation time $\tau_\mathrm{ini}$, and the condensate-formation time 
$\tau_\mathrm{eq}$ -- implies that these are set by the inverse interaction energies  \cite{kss92,svi91}, and not by the inverse cross sections \cite{sto91}, which would yield a $1/a^2$-scaling: A larger interaction energy causes faster initiation of the (quasi-)condensate formation, and also a more rapid thermalization towards the equilibrium value of the condensate fraction.
The $1/a$- rather than $1/a^2$-dependence is due to the emerging coherence between the highly occupied IR modes \cite{dwg17}.

The nonequilibrium system shows self-similar scaling \cite{gli21}, with a net flow of particles towards the infrared (\textit{bottom-up}) and into the condensed state. This is accompanied by a corresponding energy flow towards the ultraviolet carried by a relatively small number of atoms that build up the MB tail, see  Figs.\,\ref{fig1} and \ref{fig2}.



\section{Conclusions}
While statistical equilibrium for bosons and fermions is uniquely determined by temperature $T$ and chemical potential $\mu$ in the Bose--Einstein and Fermi--Dirac distributions, 
we can account for the time-dependent approach to equilibrium via a nonlinear diffusion equation.

In the approximate case of constant diffusion and drift coefficients, we have solved the equation exactly through a suitable nonlinear transformation. The nonequilibrium system behaviour is determined through
diffusion coefficient $D$ and drift coefficient $v$, the initial and final temperatures, and the initial chemical potential.
Particle-number conservation is ensured during the thermalization process via elastic two-body collisions. This is essential for the proper calculation of the time-dependent condensate fraction.

We have applied this model to the thermalization of ultracold $^{39}$K atoms with tunable interaction strength, and compared to recent Cambridge data. As in our earlier comparison with $^{23}$Na MIT data for the time-dependent condensate fraction, overall agreement with the experimental results is found. Minor deviations in the short-time behaviour at small scattering lengths might indicate the limits of a description that relies on constant transport coefficients in order to obtain an analytical solution of the problem.

The transport equation does not consider phases and -- as other Boltzmann-type numerical approaches -- does not explicitly treat the buildup of coherence in the condensate.
Nevertheless, the nonequilibrium-statistical method together with the condition of particle-number conservation allows to properly account for the time-dependent number of particles that move into the condensed state, because it includes the condition that condensate formation starts at the point in time where the chemical potential vanishes.

In agreement with the Cambridge data, the conden-\\sate-initiation time $\tau_\mathrm{ini}$ and the condensate-formation time $\tau_\mathrm{eq}$ are inversely proportional to the tunable scattering length, inducing shorter time constants
for larger scattering lengths and a universal dependence of the {(quasi-)}condensate fraction on a time parameter that scales linearly with the scattering length. The condensate-initiation time is found to be always considerably shorter than the condensate-formation time.

It would be valuable to have precise experimental information about the time-dependent condensate formation in other cold-atom systems such as $^{87}$Rb and $^7$Li using a single-quench technique to remove the high-energy atoms under similarly controlled conditions in order to test the available numerical and analytical non\-equilibrium-statistical approaches in detailed comparisons with data.
\begin{acknowledgement}
We thank Thomas Gasenzer of Heidelberg University and Zoran Hadzibabic of Cambridge University for discussions of theoretical aspects, and the experimental results.
 One of the authors (GW) is grateful to Emiko Hiyama for her hospitality at Tohoku University, Sendai, and RIKEN, Tokyo, where this work was finalized with the support of JSPS-BRIDGE fellowship BR200102.
\end{acknowledgement}
\begin{itemize}
\item Availability of data and materials:
The experimental data that are compared with our model calculations in Fig. 3 are provided in Ref. [1] through the Apollo repository\\
\small{https://www.repository.cam.ac.uk/handle/1810/315777.}\\
\normalsize
The calculated model results shown in Figs. 1-3\\
 (Mathematica-plots) are available as tables from the
corresponding author upon reasonable request.
\item Code availability: 
The Mathematica-codes for the analytical calculations shown in Figs.1, 2, and the \texttt{C++} code for the numerical
integration in Eq. (19) can be made available upon request.
\item Authors' contributions:
A. Kabelac: Numerical integration of Eq. (19) using \texttt{C++}, preparation of data for Fig.3, participation in derivations. G. Wolschin: Concept, Methodology, Writing - original draft, Supervision, Formulation and analytical solution of the NBDE, Figs. 1–3, Writing - review and editing.
\end{itemize}


\begin{thebibliography}{21}

\bibitem{gli21}
J.A.P. Glidden, C.~Eigen, L.H. Dogra, T.A. Hilker, R.P. Smith, Z.~Hadzibabic,
  \emph{Bidirectional dynamic scaling in an isolated {Bose} gas far from
  equilibrium}, Nature Phys. \textbf{17}, 457--461 (2021)

\bibitem{gz97}
C.W. Gardiner, P.~Zoller, R.J. Ballagh, M.J. Davis, \emph{Kinetics of
  {Bose-Einstein} condensation in a trap}, Phys. Rev. Lett. \textbf{79},
  1793--1796 (1997)

\bibitem{bzs00}
M.J. Bijlsma, E.~Zaremba, H.T.C. Stoof, \emph{Condensate growth in trapped
  {Bose} gases}, Phys. Rev. A \textbf{62}, 063609 (2000)

\bibitem{miesner_bosonic_1998}
H.J. Miesner, D.M. Stamper-Kurn, M.R. Andrews, D.S. Durfee, S.~Inouye,
  W.~Ketterle, \emph{Bosonic stimulation in the formation of a
  {Bose}-{Einstein} condensate}, Science \textbf{279}, 1005--1007 (1998)

\bibitem{kdg02}
M.~Köhl, M.J. Davis, C.~Gardiner, T.~Hänsch, T.~Esslinger, \emph{Growth of{
  Bose-Einstein} condensates from thermal vapor}, Phys. Rev. Lett. \textbf{88},
  080402 (2002)

\bibitem{gw18}
G.~Wolschin, \emph{Equilibration in finite {Bose} systems}, Physica A
  \textbf{499}, 1--10 (2018)

\bibitem{gw18a}
G.~Wolschin, \emph{An exactly solvable model for equilibration in bosonic
  systems}, EPL \textbf{123}, 20009 (2018)

\bibitem{gw20}
G.~Wolschin, \emph{Time-dependent entropy of a cooling {{Bose}} gas}, EPL
  \textbf{129}, 40006 (2020)

\bibitem{rgw20}
N.~Rasch, G.~Wolschin, \emph{Solving a nonlinear analytical model for bosonic
  equilibration}, Phys. Open \textbf{2}, 100013 (2020)

\bibitem{sgw21}
A.~Simon, G.~Wolschin, \emph{Time-dependent condensate fraction in an
  analytical model}, Physica A \textbf{573}, 125930 (2021)

\bibitem{snowo89}
D.W. Snoke, J.P. Wolfe, \emph{Population dynamics of a {Bose} gas near
  saturation}, Phys. Rev. B \textbf{39}, 4030--4037 (1989)

\bibitem{kss92}
Y.M. Kagan, B.V. Svistunov, G.V. Shlyapnikov, \emph{Kinetics of {Bose}
  condensation in an interacting {Bose} gas}, Sov. Phys. JETP \textbf{74},
  279--285 (1992)

\bibitem{setk95}
D.V. Semikoz, I.I. Tkachev, \emph{Kinetics of {Bose} condensation}, Phys. Rev.
  Lett. \textbf{74}, 3093--3097 (1995)

\bibitem{lrw96}
O.J. Luiten, M.W. Reynolds, J.T.M. Walraven, \emph{Kinetic theory of the
  evaporative cooling of a trapped gas}, Phys. Rev. A \textbf{53}, 381--389
  (1996)

\bibitem{hwc97}
M.~Holland, J.~Williams, J.~Cooper, \emph{{Bose-Einstein} condensation: Kinetic
  evolution obtained from simulated trajectories}, Phys. Rev. A \textbf{55},
  3670--3677 (1997)

\bibitem{jgz97}
D.~Jaksch, C.W. Gardiner, P.~Zoller, \emph{Quantum kinetic theory. {II.
  Simulation} of the quantum {Boltzmann} master equation}, Phys. Rev. A
  \textbf{56}, 575--586 (1997)

\bibitem{gw22}
G.~Wolschin, \emph{Nonlinear diffusion of fermions and bosons}, EPL
  \textbf{140}, 40002 (2022)

\bibitem{fa08}
S.~Falke, H.~Kn\"ockel, J.~Friebe, M.~Riedmann, E.~Tiemann, C.~Lisdat,
  \emph{Potassium ground-state scattering parameters and {Born-Oppenheimer}
  potentials from molecular spectroscopy}, Phys. Rev. A \textbf{78}, 012503
  (2008)

\bibitem{svi91}
B.V. Svistunov, \emph{Highly nonequilibrium {Bose} condensation in a weakly
  interacting gas}, J. Mosc. Phys. Soc. \textbf{1}, 373--390 (1991)

\bibitem{sto91}
H.T.C. Stoof, \emph{Formation of the condensate in a dilute {Bose} gas}, Phys.
  Rev. Lett. \textbf{66}, 3148--3151 (1991)

\bibitem{dwg17}
M.J. Davis, T.M. Wright, T.~Gasenzer, S.A. Gardiner, N.P. Proukakis,
  \emph{Formation of {Bose–Einstein} condensates}, in \textit{Universal
  themes of {Bose-Einstein} condensation}, Cambridge University Press (eds. N.
  P. Proukakis et al.)  (2017)

\end{thebibliography}
%
\end{document}